\documentclass[a4paper,12pt]{article}
\usepackage[cp1251]{inputenc}
\usepackage[russian]{babel}

\newcommand{\be}{\begin{equation}}
\newcommand{\ee}{\end{equation}}
\begin{document}

{\Large\bf Virial estimator for dark energy}

\vspace{1cm}

\noindent { A.D.~Chernin$^{1,2}$, P.~Teerikorpi$^{2}$,
M.J.~Valtonen$^{2}$,\\
V.P. Dolgachev$^1$, L.M. Domozhilova$^1$, G.G.~Byrd$^{3}$}

\vspace{1cm} {\it $^{1}$Sternberg Astronomical Institute, Moscow
University, Moscow, 119899, Russia,

$^{2}$Tuorla Observatory, Turku University, Piikki\"o, 21 500,
Finland}

$^{3}$Alabama University, Tuscaloosa, USA

\vspace{1cm}

{\bf \noindent A new estimator of the local density of dark energy
is suggested which comes from the virial theorem for
non-relativistic gravitating systems embedded in the uniform dark
energy background.}

\vspace{1cm}

\section{Introduction}

The virial theorem of Newtonian mechanics plays a central part in
the studies of gravitationally bound quasi-stationary astronomical
many-body systems like star clusters, galaxies, their groups and
clusters. The theorem is widely used for the mass determination of
the systems. Most impressive examples of this kind are the
discovery of dark matter in the Coma cluster by Zwicky [1] and the
discovery of dark haloes of galaxies by Einasto et al.[2]. In this
note we show that the virial theorem can also be used for
detection and measurement of dark energy, or Einstein's
cosmological constant $\Lambda$.

\section{Dark energy in terms of Newtonian mechanics}

The current standard cosmological $\Lambda$CDM model treats dark
energy macroscopically as a vacuum-like medium with density
$\rho_{\Lambda, glob} = 8 \pi G \Lambda > 0$ (where $G$ is the
Newton gravitational constant). The density is perfectly uniform
and its value is the same in any reference frame. Therefore
$\rho_{\Lambda}$ is the only physical parameter of the medium. In
the $\Lambda$CDM cosmology, the antigravity produced by dark
energy controls the global cosmological expansion on the scales
which are larger than the size of the cosmic sell of uniformity
($\simeq 300-1000$ Mpc). Though the isotropic $\Lambda$CDM model
does not extend to local volumes which are located deeply inside
the cosmic cell of uniformity, the interpretation of dark energy
adopted in the model may reasonably be extrapolated as well to
studies of local effects of dark energy. Under this assumption, we
have demonstrated earlier that dark energy dominates the dynamics
of local flows of expansion observed around nearby groups of
galaxies and the nearest Virgo cluster [3-12] on the scales 1-20
Mpc. It was also found that the local dark energy density
$\rho_{\Lambda}$ is nearly or even exactly coincides with the
global value $\rho_{\Lambda glob}$. Following this approach here,
we will nevertheless discriminate, generally, between the local
$\rho_{\Lambda}$ and global $\rho_{\Lambda glob}$ values of dark
energy.

Dark matter and cosmic baryonic matter are non-relativistic and
can  be described with the use of the classic Newtonian mechanics.
As for dark energy, it is an essentially relativistic substance,
whatever its physical nature  and microscopical structure might
be; it can adequately be described only in terms of general
relativity. It is well-known nevertheless that dark energy may be
put in the context of Newtonian mechanics (as it is explained, for
instance, in [3] and Sec. 3.4 of [4]; see also [13,14]). As a
medium dark energy is characterized by positive density,
$\rho_{\Lambda}$ and pressure $p_{\Lambda}$ which is negative.
They are related to each other by the equation of state,
$\rho_{\Lambda} = - p_{\Lambda}$ (hereafter the speed of light $c
= 1$). The medium with that equation of state is vacuum, in terms
of mechanics since rest and motion are not discriminated relative
to this medium.

According to general relativity, the source of gravity is not
matter density $\rho$ alon (as it is in the Newtonian mechanics),
but energy-momentum tensor which includes both density $\rho$ and
pressure $p$. As a result, an "effective gravitating density"
$\rho_{eff}$ is introduced which is simply the sum $ \rho + 3p$,
if the medium is uniform. This may directly be seen in the
Friedmann ("first") equation, as well as in the de Sitter solution
and some other general relativity relations. For dark energy with
its equation of state $\rho_{\Lambda} = - p_{\Lambda}$ one has:
\be \rho_{eff} = \rho_{\Lambda} + 3p_{\Lambda}= - 2\rho_{\Lambda}
< 0.\ee \noindent Negative effective density means that dark
energy produces a repulsion force, or antigravity.

This result is borrowed from general relativity to Newtonian
mechanics (see in Secs.2.1 and 3.4 of [4]). Consequently, in
Newtonian description, dark energy antigravity is represented by
the repulsive antigravity force (per unit mass)
\be {\bf F}_{\Lambda} = + {\frac{8 \pi}{3}} G \rho_{\Lambda} {\bf
r}, \ee

where the radius-vector ${\bf r}$ denotes the spatial position of
a particle. The corresponding potential
\be \Phi_{\Lambda} = - {\frac{4 \pi}{3}} G \rho_{\Lambda} r^2. \ee

The relations (2),(3) come directly from the Schwarzschild-de
Sitter spacetime considered in the limit of the weak deviations
from the Galilean metric (see again section 3.4 of [4]). This
approximation is quite appropriate to the physical conditions in
groups and clusters of galaxies on the spatial scales 1-20 Mpc we
have here in mind, since |$\Phi_{\Lambda}|/c^2 \sim (0.3-1) \times
10^{-6} << 1$. Because of the same reason, the local curvature
effects are considered negligible. Note also that the general
cosmological expansion does not affect the dynamics of local
quasi-stationary systems like groups and clusters; because of this
the Friedmann cosmological equations are not involved in the
description of local dark energy effects.

General relativity indicates also that the inertial and passive
gravitational mass of dark energy (the both are $\rho_{\Lambda} +
p_{\lambda}$, per unit mass) are zero; it means in particular that
dark energy is not affected by its own antigravity and therefore
its self-gravity potential energy is zero (which is important for
considerations of the next section).

\section{Modified virial theorem}

The first versions of the virial theorem with $\Lambda$ were
proposed decades ago [15,16]; more recently various aspects of the
modified theorem have been discussed in [12,17-20]. Here we show
briefly how the theorem may be deduced with the use of Eqs.(2),
(3).

As is long known, a finite collection of point-mass particles
interacting gravitationally via the Newtonian potential $\Phi_M
\propto 1/r$ and bounded in space obeys the virial relation $<K> =
-{\frac{1}{2}} <U>$, where $<K>$ is the time average of the total
kinetic energy of the system, and $<U>$ is the time average of its
total potential energy. Generally, the virial theorem states that
if the potential is a homogeneous function of order $n$ of the
coordinates, then the average kinetic energy $<K>$ and average
potential energy $<U>$ of the system relate to each other as $<K>
= {\frac{n}{2}} <U>$.

If a system of $N$ gravitating particles is embedded in the
uniform dark energy background, the particles move in the
potential which is the sum of two components: the Newtonian one
$\Phi_M$ with $n = -1$ for the particle-particle interaction and
the dark energy potential $\Phi_{\Lambda}$ with $n=2$ (see Eq.(3))
for the particle-dark energy interaction. Therefore one may expect
that a modified virial theorem for the sum of the potentials must
include two items: the conventional  $- {\frac{1}{2}} <U>$ and a
new one $+ <W>$, where $W$ is the potential energy of the
particles which is due to their interaction with the dark energy
background (see below).

Let us remind (see references in Sec.2) that the generalized
virial relation that accounts for both matter gravity and dark
energy antigravity can directly be obtained from the Newtonian
mechanics of an arbitrary bound $N$-body system with masses $m_i$,
where $i = 1,2,3,...,N$. The potential energy of the interaction
between the $i$-th and the $j$-particles
\be U_{ij} = - \frac{G m_i m_j}{r_{ij}}, \ee \noindent where ${\bf
r_{ij} = r_i - r_j}$. The particle-particle contribution to the
total potential energy of the system is the sum over all N
particles:
\be U = - {\frac{1}{2}} \sum_i^N \sum_{j \ne i}^N \frac{G m_i
m_j}{r_{ij}}. \ee

The potential energy of the interaction of the $i$-th particle
with the dark energy background (see Eq.(3))
\be W_i =  m_i \Phi_{\Lambda} = - m_i {\frac{4 \pi}{3}} G
\rho_{\Lambda} r^2. \ee The contribution of the particle-dark
energy interaction to the total potential energy
\be W = - \sum_i^N m_i {\frac{4 \pi}{3}} G \rho_{\Lambda} r^2. \ee

According to Eq.(2), the equation of motion for the $i$-th
particle moving with a non-relativistic velocity
\be m_i {\bf \ddot  r_i} = - \sum_{j \ne i}^N {\frac{G m_i
m_j}{r_{ij}^3} (\bf r_i - \bf r_j)} +  2 m_i {\frac{4 \pi}{3}} G
\rho_{\Lambda} {\bf r_i}. \ee

The scalar product of both sides of this equation by $\bf r$
\be m_i ({\bf r_i \ddot r_i }) = - \sum_{j \ne i}^N {\frac{G m_i
m_j}{r_{ij}^3} (\bf r_i - \bf r_j ) \bf r_i} +  2 m_i {\frac{4
\pi}{3}} G \rho_{\Lambda} {\bf r_i}^2. \ee

Having in mind that the particle velocity $\bf v_i = {\bf \dot
r_i}$ and also
\be {\frac{d}{dt}}({\bf v_i r_i}) = {\bf v_i}^2 + {\bf \dot v_i
r_i}, \ee we rewrite the left side of Eq.(9):
\be m_i {\frac{d}{dt}}({\bf v_i \bf r_i} ) - 2 K_i, \ee

\noindent where $K_i = {\frac{1}{2}} m_i v^2$ is the kinetic
energy of the particle. The sum over $i$ and a simple identical
transformation of Eq.(9) (about the transformation of the first
item in the right side see, for example, [21]) lead to
\be \frac{1}{2} m_i {\frac{d}{dt}}({\bf v_i r_i}) - K =
\frac{1}{2} U - W, \ee

The potential energies $U$ and $W$ are given by Eqs.(5),(7). The
average over a large time of the first item in the left side here
is zero, if the particles are located in a bound finite volume.
Then we finally find:
\be K = -{\frac{1}{2}} U +  W. \ee

This is the modified virial theorem for a N-body quasi-stationary
bound system embedded in the dark energy background. The result
shows how the classic relation which treats each of the potential
energies separately (see above) reveals itself in real
astronomical systems: the virial relation contains two items in
the right side, the first of them (positive) is conventional, and
the second one (negative) is due to the dynamical effects of dark
energy.

\section{Theorem for one particle}

The generalized virial relation of Eq.(13) may be derived for one
light particle of mass $m$ moving in the central field of a heavy
mass $M$ along a finite bound orbit in the dark energy background.
The mass $M$ produces the potential $\Phi_M = - GM/r$ and the dark
energy background produces (as above) the potential
$\Phi_{\Lambda} = - {\frac{4 \pi}{3}} G \rho_{\Lambda} {\bf r}^2.$
The particle equation of motion
\be {\bf \dot v} = - {\frac{GM}{r^2}} {\frac{{\bf r}}{r}} +
{\frac{8 \pi}{3}} G \rho_{\Lambda} {\bf r}. \ee

Multiplying both sides of Eq.(14) by $\bf r$, one gets:
\be {\bf \dot v} {\bf r}   = - {\frac{GM}{r}}  +  {\frac{8
\pi}{3}} G \rho_{\Lambda} {r}^2 = \Phi_M - 2 \Phi_{\Lambda}. \ee

The transformation of Eq.(10) applied to Eq.(15) leads now to
\be {\frac{d}{dt}}({\bf v r}) = v^2 - {\frac{GM}{r}} + {\frac{8
\pi}{3}} G \rho_{\Lambda} {r}^2 = v^2 + \Phi_M - 2 \Phi_{\Lambda}.
\ee

Or in terms of energies $ K = {\frac{1}{2}} mv^2, U = m \Phi_M, W
= m \Phi_{\Lambda}$:
\be {\frac{1}{2}}{\frac{d}{dt}}({\bf v r}) = K + {\frac{1}{2}} U -
W. \ee

Averaging over time gives again the virial relation in form of
Eq.(13) which is valid now for the finite motion of one particle
as well as for the bound N-body system.

The particle potential energy  $W$ due to dark energy is an
essentially negative value. Because of this, the kinetic energy of
a virialized system is less in the presence of the dark energy
background than in the perfectly empty space. The physical cause
is clear: the dark energy antigravity cancels partly the matter
gravity; as a result, the potential well of the system is not as
deep as it would be in empty space. Accordingly, the velocities of
the particles are less in this case. Since the kinetic energy is
positive, Eq.(13) implies that
\be |U|  > 2 |W|. \ee

This inequality is the necessary condition for the existence of a
system with finite bound orbits. Eq.(18) guaranties that gravity
is stronger than antigravity in the volume of the system. Indeed,
if a steady-state characteristic radius of a system is $r$, and
its total mass is $M$, then $|U| = GM^2/r$ and $ 2 |W| =  M
{\frac{8 \pi}{3}} G \rho_{\Lambda} r^2$. Then Eqs.(13),(18) give:
\be r < R_{\Lambda} = [{\frac{3M}{8 \pi \rho_{\Lambda}}}]^{1/3}
\simeq 1 \times (M_{12})^{1/3} Mpc.\ee

Here $R_{\Lambda}$ is the "zero-gravity radius" [6,7]: at the
distance  $r = R_{\Lambda}$ the gravity and antigravity of the
system completely balance each other.  In Eq.(19), $M_{12}$ is the
mass in the units of $10^{12} M_{\odot}$, and the observed value
$\rho_{\Lambda} = 7.5 \times 10^{-30}$ г см$^{-3}$ is used.

Inequality (19) has a rather universal sense: the size of any
gravitationally bound system must obey it in the world of dark
energy.

\section{Virial estimators}

The relation of Eq.(13) provides a mass estimator which
generalizes the conventional virial estimator:
\be M = {\frac{v^2 r}{G}} + {\frac{8 \pi}{3}} \rho_{\Lambda}
r^3,\ee

\noindent where $r$ and $v$ are characteristic size and velocity
of the system. Eq.(20) shows that the mass of astronomical system
is systematically underestimated, if dark energy is not taken into
account.

The effect of dark energy is given by the new positive item in the
right side of Eq.(20). The term is quantitatively equivalent to
the effective mass of dark energy contained in the volume of the
system. The physical sense of the extra term is clear. Indeed, the
antigravity of dark energy cancels partly the gravity of dark
matter and baryons in the volume of the system. Consequently, the
characteristic velocity $v$ of the first term in the right side of
Eq.20 represents the partly cancelled gravity of dark matter and
baryons, not their total gravity. The second term in the right
side of Eq.(20)  "restores" the true mass of the dark matter and
baryons. The mass correction is given by the relation $M_{vir} =
{\frac{v^2 r}{G}}(1 + f)$, where $f$ is the ratio of the new item
to the conventional one:
\be f \equiv {\frac{8 \pi}{3}} G \rho_{\Lambda}(r/v)^2 =
(\frac{t_{cross}}{t_{\Lambda}})^2. \ee

\noindent Here $t_{cross} = r/v$ is the system crossing time;
$t_{\Lambda} = ({\frac{8 \pi G}{3}} G \rho_{\Lambda})^{-1/2} = 16
$ Gyr is the dark energy characteristic time which is near the
current age of the Universe. The ratio $f$ is less than unity for
any gravitationally bound system:
\be f = (\frac{t_{cross}}{t_{\Lambda}})^2 \simeq 0.4 ({\frac{r}{1
Mpc}})^2 ({\frac{v}{100 km/s}})^{-2}. \ee

Having in mind all the necessary reservations regarding a degree
of virialization of a system, its age, shape, internal structure,
the way in which available data are obtained, etc., we may
illustrate the dark energy effect in Eq.(20) by simplest examples.
For a halo of giant galaxy like the Milky Way or M31 with $r
\simeq 0.3$ Mpc, $v \simeq 200$ km s$^{-1}$, so that the ratio $f
\simeq 0.03$; for a typical sparse group of galaxies like the
Local Group with $r \simeq 1$ Mpc, $v \simeq 70$ km s$^{-1}$, the
ratio $f \simeq 0.8$, and for a big cluster of galaxies with $r
\simeq 5$ Mpc, $v \simeq 1 000$ km s$^{-1}$, the ratio $f \simeq
0.04$. These figures show that the effect of dark energy is most
prominent for groups of galaxies: the correctly estimated mass is
30-50\% larger than that made with the traditional virial
estimator.

Eq.(20) shows that
\be \rho_{\Lambda} =  \frac{3}{8\pi G} [ \frac{GM}{r^3} -
\frac{v^2}{r^2} ].\ee

\noindent If the matter mass $M$ of a system, its characteristic
size $r$ and velocity $v$ are independently measured, the system
may serve as a measurement setup for local (on the spatial scale
of the system) dark energy detection, and Eq.(23) turns out to be
an estimator of the density $\rho_{\Lambda}$. In terms of the
crossing time $t_{cross}$, cosmological critical density $\rho_c =
\frac{3}{8\pi G}H^2$ and the Hubble factor $H$ the estimator takes
the form:
\be \rho_{\Lambda} =  \frac{1}{2} <\rho> - \rho_c
\frac{1}{(Ht_{cross})^2},\ee

\noindent where $<\rho> = \frac{M}{\frac{4\pi}{3}r^3}$ is the mean
density of the system. Or in the units of the critical density:
\be \Omega_{\Lambda} = \frac{1}{2}<\Omega> - (Ht_{cross})^{-2},
\ee

\noindent where $<\Omega> = \frac{<\rho>}{\rho_c}$.

For a galaxy group with the mass $M = (1-2)\times 10^{12}
M_{\odot}$, the size $r = 1$ Mpc and the velocity $v = 70$ km/s,
we have $<\Omega> = 3-6$ and the product $Ht_{cross} \simeq 1$.
Then the estimator of Eq.(25) gives:
\be \Omega_{\Lambda} = 0.5-2. \ee

\noindent The interval  of possible values is not too wide in
Eq.(26), but it contains comfortably the value $\Omega_{\Lambda
glob} = 0.70-0.75$ which is known from the global cosmological
observations [22]. The result implies that the local density of
dark energy is near the global value, if not coincides with it
exactly, -- in complete agreement with our earlier findings (see
Sec.2).

\section{Conclusions}

To summarize, the modified virial theorem provides us with a
modified mass estimator and -- which is perhaps more interesting
-- a new estimator for the dark energy density on the local scale
of groups and clusters of galaxies. The virial mass determination
has proved its usefulness in decades of studies. However practical
effectiveness of the virial estimator for dark energy is not yet
so obvious; the theoretical recipe we suggest for virial detection
and measurement of local dark energy needs more empirical studies.

{\bf Acknowledgements.} A.C., V.D., and L.D. appreciate a partial
support from the RFBR grant 02-10-00178.

\vspace{0.5cm}

\section*{References}

\small

1. F. Zwicky, Helv. Phys. Acta {\bf 6}, 110 (1933)

2. J. Einasto, A. Kaasik, E. Saar, Nature {\bf 250}, 309 (1974)

3. A.D.Chernin, Physics-Uspekhi {\bf 44}, 1099 (2001)

4. A.D.Chernin, Physics-Uspekhi {\bf 51}, 253 (2008)

5. A.D.Chernin, P.Teerikorpi, Yu.Baryshev Yu.V.,
[astro-ph/0012021] Adv. Space Rev. {\bf 31}, 459 (2003)

6. G.G. Byrd, A.D. Chernin, M.J. Valtonen, Cosmology: Foundations
and Frontiers (Editorial URRS, Moscow 2007)

7. A.D. Chernin, I.D. Karachentsev, M.J. Valtonen, et al.
Astron.Astrophys. {\bf 415}, 19 (2004)

8. P. Teerikorpi, A.D. Chernin, Yu.V. Baryshev, Astron.Astrophys.
{\bf 440}, 791 (2005)

9. P. Teerikorpi, A.D. Chernin, I.D.Karachentsev, M.J., Valtonen
Astron.Astrophys. {\bf 483}, 383 (2008)

10. A.D. Chernin, I.D. Karachentsev, P. Teerikorpi P., et al.,
Grav.Cosm. {\bf 16},1 (2010)

11. I.D. Karachentsev, Nasonova O.G., MNRAS 405, 1075 (2010)

12. A.D. Chernin, I.D. Karachentsev, O.G. Nasonova, et al.,
Astron. Astrophys. 520, A104 (2010)

13. E.J. Copeland, M. Sami, S. Tsujikava, IJMPD 15, 1753 (2006)

14. M. Sami 2009, ArXiv: 0904.3445

15. W.R. Forman, Astrophys.J. 159, 719 (1970)

16.  J.C. Jackson, MNRAS 148, 249 (1970)

17.  O. Lahave et al., MNRAS {\bf 251}, 128 (1991)

18. M. Nowakowsky, J.-C. Sanabria, A. Garcia Phys.Rev. {\bf
D}66:023003 (2002)

19.  A.D. Chernin, V.P. Dolgachev, L.M. Domozhilova, et al.,
Astron. Rep. 54, 185 (2010)

20.  G.S. Bisnovatyi-Kogan, M.Merafiva, S.O. Tarasov, arXiv
1102.0972 (2011)

21.  C. Kittel, M.D. Knight, M.F., Ruderman. Berkeley Physics
Course: Mechanics (Mcgraw-Hill Coll., New York 1965)

22. D.N. Spergel et al., ApJS  {\bf 170}, 337 (2007)


\end{document}